\documentclass[epj,referee]{svjour}
\usepackage{graphics}
\begin{document}
\title{Direct experimental text of non-separability and other quantum techniques using continuous variables of light}
\author{N. Korolkova, Ch. Silberhorn, O. Gl\"ockl,
S. Lorenz, Ch. Marquardt, G. Leuchs} 
%
%
\institute{Zentrum f\"ur Moderne Optik an  der Universit\"at
Erlangen-N\"urnberg,  Staudtstra{\ss}e 7/B2, D-91058 Erlangen,
Germany. }
\date{Received: date / Revised version: date}
%
\abstract{ We present schemes for the generation and evaluation of continuous variable
entanglement of bright optical beams and give a brief overview of a variety of optical
techniques and quantum communication applications on this basis. A new entanglement-based
quantum interferometry scheme with bright beams is suggested. The performance of the
presented schemes is independent of the relative interference phase which is advantageous
for quantum communication applications.
} 
\maketitle
\section{Introduction}
\label{intro} Quantum information processing using continuous
variables \cite{BraunsteinLloyd,BraunsteinKimble,Reid} has emerged
as an alternative approach to quantum information with single
photons and atoms described with discrete variables. The main
motivation for this new approach is the availability of controlled
sources, efficient detection systems and easy-to-handle processing
using linear elements. The price  one has to pay for these
advantages is always non-maximal entanglement. What is continuous
variable entanglement about? We first give  a short introduction
providing an intuitive picture of how continuous variable
entanglement manifests itself. Consider two spatially separated
optical modes $j = 1, 2$. The involved optical fields can be fully
described by means of field quadratures \cite{MilburnWalls}, the
amplitude quadrature $\hat X_j = \hat a^\dagger_j + \hat a_j$ and
the phase quadrature $\hat Y_j = i (\hat a^\dagger_j - \hat a_j)$.
They obey the following commutation rules:
\begin{eqnarray}
 \left [ \hat X_j, \hat Y_k \right ] =
i \delta_{jk}, \  j, k = 1, 2.
\end{eqnarray}
These commutation relations leave the possibility for  combined
variables of both modes to commute:
\begin{eqnarray}
 \left [ \hat X_1 + \hat X_2, \hat Y_1 - \hat Y_2 \right ] = 0.
\end{eqnarray}
Hence quantum states are possible, for which all the variables
$\hat X_j, \hat Y_j$ are uncertain, but certain joint variables of
two optical modes together are both well defined:
\begin{eqnarray}
 \hat X_1 + \hat X_2 \rightarrow {\rm well\  defined}, \nonumber \\
 \hat Y_1 - \hat Y_2 \rightarrow {\rm well\  defined}.
\end{eqnarray}
 In the quantum optical context this is known as two-mode squeezing
for $\hat X$ or $\hat Y$ \cite{MilburnWalls}. The measure for the
degree of two-mode squeezing is the variances of the sum or
difference signal of two modes, which approache zero for perfect
squeezing:
\begin{eqnarray}
 V(\hat X_1 + \hat X_2) \rightarrow 0, \nonumber \\
 V(\hat Y_1 - \hat Y_2) \rightarrow 0, \label{TwoModeSq}
\end{eqnarray}
where $V(A)=\langle \hat A^2 \rangle - \langle \hat A \rangle^2$ denotes a variance of an
operator $\hat A$.  Equations (\ref{TwoModeSq}) also represent a measure for quantum
correlations between two spatially separated optical modes, i.e. for continuous variable
entanglement.
\section{Experimental evaluation of continuous variable entanglement}
\label{secExpEv}  Can we indeed associate such kind of
correlations with entanglement? An entangled state  is a
non-separable quantum state. It means that the state of a system
cannot be represented as a convex sum of product states of two
subsystems. The necessary and sufficient condition for
separability of a system for discrete variables is given by the
Peres-Horodecki criterion \cite{PeresHorodecki}. This condition
has been recently extended to continuous variable two mode
Gaussian states \cite{Duan,Simon}, like two mode squeezed states
(\ref{TwoModeSq}). The criterion derived by Duan {\it et al} has
an attractive potential for experimental quantum communication  as
it can be expressed in terms of  observable quantities. They can
then be  measured using the conventional toolbox of experimental
quantum optics.  In this spirit we re-express the results of
\cite{Duan,Simon} in terms of the amplitude $\hat X_j = \hat
a^\dagger_j + \hat a_j $ and the phase $\hat Y_j = i (\hat
a^\dagger_j - \hat a_j)$ quadratures of bright beams $j=1,2$.
Because of the high intensity of the optical fields involved, we
use the linearization approach throughout the paper: $\hat X_{\rm
j} = \langle X_{\rm j} \rangle + \delta \hat X_{\rm j}$, $\hat
Y_{\rm j} = \langle Y_{\rm j} \rangle + \delta \hat Y_{\rm j}$.
The entangled quantities are then the quantum uncertainties in the
respective field quadratures. The non-separability criterion for
the quantum state of two optical modes requires:
\begin{eqnarray}
\displaystyle
   V_{\rm sq}^\pm (\delta X) + V_{\rm sq}^\mp(\delta Y)
 < 2, \label{pereshor}
    \end{eqnarray}
\begin{eqnarray}
\displaystyle
   V_{\rm sq}^\pm (\delta X) = \frac{V(\delta \hat{X}_{1}
 \pm   g\ \delta \hat{X}_{2})}{V( \hat{X}_{1, {\rm SN}} +
   g \  \hat{X}_{2, {\rm SN}})}, \label{sq-amp}
    \end{eqnarray}
        \begin{eqnarray}
   V_{\rm sq}^\mp(\delta Y) = \frac{V(\delta \hat{Y}_{1}
   \mp  g\ \delta \hat{Y}_{2})}{V( \hat{Y}_{1, {\rm SN}}+ g\
\hat{Y}_{2, {\rm SN}})}
   \label{sq-ph}
\end{eqnarray}
where $g$ is a variable gain. The variances labeled "SN" correspond to the shot noise
level of respective beams which marks a boundary between classical and quantum regime.
The upper (lower) signs hold for the anti-correlated (correlated) amplitude quadratures
and correlated (anti-correlated) phase quadratures. The squeezing variances in Eq.
(\ref{pereshor}, \ref{sq-amp}, \ref{sq-ph}) are the normalized variances of Eq.
(\ref{TwoModeSq}). $V_{\rm sq} = 1$ corresponds, e.g. to coherent states in both optical
modes and $V_{\rm sq} < 1$ corresponds to two mode squeezing, hence the name squeezing
variances. They are quantities measurable in an experiment and will be used throughout
this paper for the experimental evaluation of continuous variable entanglement.

\begin{figure}[t]
\begin{center}
\resizebox{0.35\textwidth}{!}{%
  \includegraphics{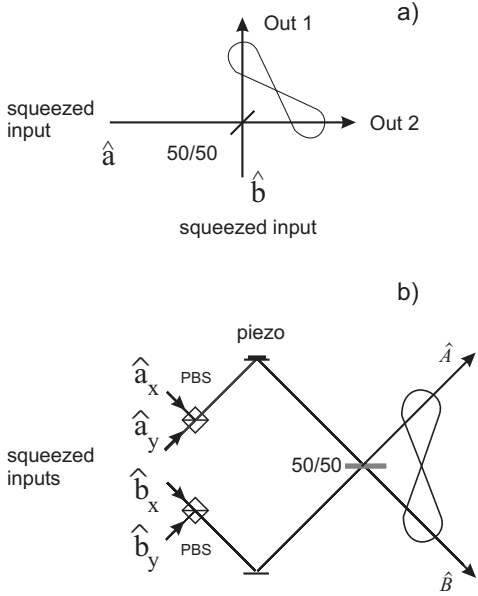}
}
\caption{Schemes for generation of continuous variable
entanglement by linear interference of squeezed beams: a)
Entanglement of amplitude and phase quadratures; b) Continuous
variable polarization entanglement. Operators $\hat a, \hat b$
describe input bright squeezed beams, indices $x, y$ denote two
orthogonal polarization modes, 50/50 is a beam splitter with a
respective splitting ratio and PBS stands for a polarization beam
splitter.}
\label{scheme}       
\end{center}
\end{figure}
\subsection{Entanglement of amplitude and phase quadratures}
\label{AmplPh} The generation of entanglement with continuous variables typically uses
optical parametric down conversion in a sub-threshold OPO. This process creates two
vacuum states with quantum correlated amplitude and anti-correlated phase quadratures or
vice versa \cite{Ou92,Furusawa98,Peng00}. Another promising scheme for the generation of
EPR-entangled beams utilizes the superposition of two independently squeezed bright light
fields to create quantum correlations  (\cite{Leonhard,Leuchs99} and Figure \ref{scheme}
(a)). For this purpose the optical phases of the incoming fields are chosen such that the
initial fields acquire a phase difference of $\frac{\pi}{2}$ for the  linear interference
at a beam splitter. This scheme was recently implemented using the Kerr-nonlinearity of
an optical fibre to produce two bright amplitude squeezed pulsed light fields
\cite{Silberhorn01}. Interference of these two  beams generates EPR-entanglement with
anti-correlated amplitude quadrature and correlated phase quadrature with $V_{\rm sq}^\pm
(\delta X) + V_{\rm sq}^\mp(\delta Y) = 0.80 \pm 0.06$ \cite{Silberhorn01}.

\subsection{Polarization entanglement}
\label{Pol} Entanglement of the amplitude and phase quadratures is familiar but of course
not the only possible type of continuous variable entanglement. Another promising set of
continuous variables are the polarization variables, the quantum Stokes operators (see
\cite{Korolkova01} and references therein).

The Stokes operators  are defined in analogy to the Stokes
parameters in classical optics:
\begin{eqnarray} \hat S_0 = \hat n_x + \hat
n_y,\label{0}\\ \hat S_1 = \hat n_x - \hat n_y,  \label{1}\\
\hat S_2
= {\hat a^\dagger}_x \hat a_y + {\hat a^\dagger}_y\hat a_x, \label{2}\\
\hat S_3 = i[{\hat a^\dagger}_y \hat a_x - {\hat a^\dagger}_x\hat
a_y] \label{3}
\end{eqnarray}
where $\hat a_k^\dagger\  (\hat a_k)$ are photon creation
(annihilation) operators and $\hat n_k= \hat a_k^\dagger \hat a_k$
is the photon number operator in polarization mode $k$  with the
basic set of the orthogonal modes linearly polarized along $x$ and
$y$ directions ($k=x,y$). The operators $\hat a_k, \hat
a_l^\dagger$ obey the usual commutation relation $[\hat a_k, {\hat
a_l}^\dagger] = \delta_{kl},  \  (k,l=x,y)$. The operators $\hat
S_j$ (j=1,2,3) satisfy the commutation relations of the Lie
algebra of the SU(2) group, for example:
\begin{eqnarray} [\hat S_2, \hat S_3] = i2 \hat S_1.
\label{StokesComm}
\end{eqnarray}
The other relations are obtained from (\ref{StokesComm}) by cyclic permutation of the
labels.

There are quantum uncertainties associated with each of the Stokes
operators (\ref{0}-\ref{3}). A quantum state of an optical field
is said to be {\it polarization squeezed}  if the uncertainty in
one of the Stokes operators is reduced below that of coherent
light, i.e. $V_j \le V^{coh}\  (\ j = 1, 2, 3)$, at the cost of
increased uncertainty in the other Stokes parameters
\cite{Korolkova01}. Such nonclassical polarization squeezed states
can be generated by superimposing two orthogonally polarized
amplitude squeezed light beams on a polarizing beam splitter
\cite{Korolkova01}.

Polarization entanglement of continuous  variables can in turn be
generated by linear interference of two polarization squeezed
beams. This is in analogy to the interference of amplitude
squeezed beams \cite{Leuchs99,Silberhorn01} for the generation of
bright EPR entanglement of the amplitude and phase quadratures
(Fig. \ref{scheme}).

For bright beams, polarization entanglement implies quantum correlations of the
uncertainties of the Stokes operators in two spatially separated optical modes $\hat A,
\hat B$ (Fig. \ref{scheme} (b)). Let us assume that polarization squeezed beams for
entanglement  generation are produced by interference of orthogonally polarized amplitude
squeezed beams with equal squeezing $V(\delta X)<1$ in each of the four input modes (Fig.
\ref{scheme} (b)).  It follows from (\ref{0}-\ref{3}, \ref{StokesComm}) that for  input
beams of equal amplitude only operators $\hat S_{1A,B}$ and $\hat S_{3A,B}$ are conjugate
variables because only $V(S_3)V(S_1) \ge  \left \vert \langle \hat S_2 \rangle \right
\vert \neq 0$. The non-separability of the quantum polarization state of two modes $\hat
A$ and $\hat B$ then means \cite{Korolkova01}:
\begin{eqnarray}
\displaystyle
   V_{\rm sq}^+ (\delta S_1)  + V_{\rm sq}^-(\delta S_3) < 2
   \label{Stokesperes}
    \end{eqnarray}
with the squeezing variances equal to:
\begin{eqnarray}
\displaystyle
   V_{\rm sq}^+ (\delta S_1) = \frac{V(\delta \hat{S}_{1A}
 +  g\  \delta \hat{S}_{1B})}{V( \hat{S}_{1A, {\rm SN}} + g\
    \hat{S}_{1B, {\rm SN}})} = V(\delta X), \label{s1}
    \end{eqnarray}
    \begin{eqnarray}
   V_{\rm sq}^-(\delta S_3) = \frac{V(\delta \hat{S}_{3A}
- g\ \delta \hat{S}_{3B})}{V( \hat{S}_{3A, {\rm SN}}+ g\
\hat{S}_{3B, {\rm SN}})} = V(\delta X).
   \label{s2}
\end{eqnarray}
The important advantage of such non-classical polarization states is that it is possibile
to measure the relevant conjugate variables  by direct detection only, using linear
optical elements.

\section{Optical techniques in direct detection}
\label{OpTechDirDet} The  use of bright beams for entanglement
generation allows the implementation of a range of optical
techniques with a particularly simple detection system. Unlike
many schemes described elsewhere, those proposed here do not
require explicit measurements of the phase quadrature of the light
field. This advantage is brought about by the use of a simple
interferometric setup which avoids more cumbersome measurements,
in particular those involving local oscillator fields. In the
following sections we describe these techniques in detail using
the entanglement of field quadratures as an example (Sec.
\ref{AmplPh} and Fig. \ref{scheme} (a)).

Prior to doing this it is worthwhile to contrast quadrature
entanglement and polarization entanglement. For polarization
entanglement (Sec. \ref{Pol} and Fig. \ref{scheme} (b)) the
explicit measurement of all the relevant conjugate variables and
of their variances  can be performed in direct detection
\cite{Korolkova01}.  For example, it is straightforward to record
$ V_{\rm sq}^+ (\delta S_1)$, $V_{\rm sq}^-(\delta S_3)$ and
calculate (\ref{Stokesperes}). It is thus not necessary to detect
the non-separability of polarization state in an indirect way. If
the schemes illustrated in the following sections were though to
use polarization entanglement, they would be more sophisticated,
requiring additional linear elements, such as half-wavelength
plates and polarizers, to record the variances of the Stokes
operators \cite{Korolkova01}.

Thus both quadrature and polarization entanglement have their
advantages and disadvantages. These determine the scope of
applications of the two different entanglement types as discussed
in Sec. \ref{AmplPh} and \ref{Pol}. Continuous variable
polarization entanglement is best suited to schemes requiring
explicit measurements of the involved conjugate variables, e.g. to
quantum cryptography. In contrast, quadrature entanglement is more
appropriate for schemes involving direct intensity detection, e.g.
quantum dense coding, quantum teleportation.

\subsection{Measuring non-separability of a quantum state}
\label{PeresDirDet} \begin{figure}[b]
\begin{center}
\resizebox{0.33\textwidth}{!}{%
  \includegraphics{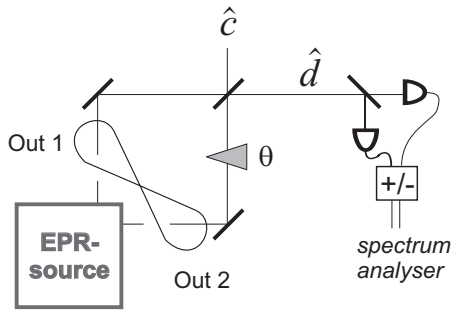}
}
\caption{Direct detection of the non-separability criterion}
\label{peres}       
\end{center}
\end{figure}
In this section we show how the
non-separability of the quantum states described in Sec.
\ref{AmplPh}, \ref{Pol} can be verified in an experiment. To
determine  the non-separability criterion of Eq. (\ref{pereshor}),
the EPR source is treated as a black box. We would like to test
now for continuous variable entanglement of the two output optical
modes  of Sec. \ref{AmplPh} and Figure \ref{scheme} (a). These two
modes are superimposed at a beam splitter (Fig. \ref{peres}). The
field operators of the beams after this interference are denoted
$\hat c$, $\hat d$. To adjust the relative interference phase, a
variable phase shift $\theta$ can be introduced in one of the
arms. The balanced detection is performed in one of the output
arms, $\hat c$ or, equivalently, $\hat d$. Difference and sum
photocurrents are recorded and the relative phase $\theta$ is
scanned. The difference photocurrent provides the shot noise level
reference:
\begin{eqnarray}
V(\delta X_{\rm c,d, SN}) =\alpha^2 \left (1 \pm \cos \theta
\right ).\label{peresminus}
\end{eqnarray}
The two signs refer to the two complementary outputs. The sum photocurrent delivers the
amplitude noise variance of the signal in the output $\hat c$ or $\hat d$:
\begin{eqnarray}
\displaystyle
 \langle \left ( \delta X_{\rm c,d} \right )^2 \rangle = \nonumber \\
 \frac{1}{2} \alpha^2\left [ \left (1 \pm \cos \theta
\right )^2 V_{\rm sq}^+ (\delta X) + \sin^2 \theta \ V_{\rm sq}^-
(\delta Y) \right ].\label{peresplus}
\end{eqnarray}
From the measurements (\ref{peresminus}, \ref{peresplus}) the
normalized noise variance can be inferred:
\begin{eqnarray}
\displaystyle V \left ( \delta X_{\rm c,d} \right ) =\frac{
\langle \left ( \delta X_{\rm c,d} \right )^2 \rangle}{V(\delta X_{\rm c,d, SN})} = \nonumber \\
\frac{1}{2} \left [  \left (1 \pm \cos \theta \right ) V_{\rm
sq}^+ (\delta X) + \frac{\sin^2 \theta} {\left (1 \pm \cos \theta
\right )} \ V_{\rm sq}^- (\delta Y) \right ]. \label{phaseDep}
\end{eqnarray}
It contains the information about both relevant squeezing
variances $V_{\rm sq}^+ (\delta X), V_{\rm sq}^- (\delta Y)$ which
characterize the quantum correlations between optical modes $1,2$
(Fig. \ref{peres}). For $\theta={\pi \over 2}$ the measured noise
variance (\ref{phaseDep}) is directly proportional to the
Peres-Horodecki criterion (\ref{pereshor}) for continuous
variables:
\begin{eqnarray}
\displaystyle V \left ( \delta X_{\rm c,d} \right ) = \frac{1}{2}
\left [  V_{\rm sq}^+ (\delta X) +  V_{\rm sq}^- (\delta Y) \right
]. \label{criterion-measured}
\end{eqnarray}
If the interfering amplitude squeezed beams are of equal squeezing, the output entangled
beams have symmetric circular uncertainty regions and $V_{\rm sq}^+ (\delta X)=V_{\rm
sq}^- (\delta Y)= V_{\rm sq}$. Then equation (\ref{phaseDep}) takes the form:
\begin{eqnarray}
\displaystyle V \left ( \delta X_{\rm c,d} \right ) =\frac{
\langle \left ( \delta X_{\rm c,d} \right )^2 \rangle}{V(\delta
X_{\rm c,d, SN})} =
 V_{\rm
sq} \label{phaseIndep}
\end{eqnarray}
and the non-separability condition reads: $V_{\rm sq} <1$ independent of the relative
interference phase $\theta$. Using (\ref{criterion-measured}) or (\ref{phaseIndep}), the
non-separability of the state can be reliably verified experimentally using the
interferometric scheme of Fig. \ref{interfer}.

\subsection{Quantum interferometry}
\label{Qinter} The interferometric detection of the
non-separability criterion provides a new and interesting insight
into the performance of the scheme (Fig. \ref{peres}) as a setup
for high-precision measurements of small phase modulations. There
are quantum limitations to the sensitivity of optical
interferometers in addition to mechanical, thermal, and other
effects. For example, the standard quantum limit for minimal
resolvable phase modulations derived from the Heisenberg
uncertainty principle reads:
\begin{eqnarray}
\displaystyle \delta \theta_{\rm min} = \sqrt{\frac{1}{n}}
\label{sql}
\end{eqnarray}
where $n$ is a total photon number in the measured mode. Quantum
interferometry attempts to exploit non-classical states of light
to go beyond this standard quantum limit in high-sensitive
measurements \cite{MilburnWalls,Interferometry}.
\begin{figure}
\begin{center}
\resizebox{0.35\textwidth}{!}{%
  \includegraphics{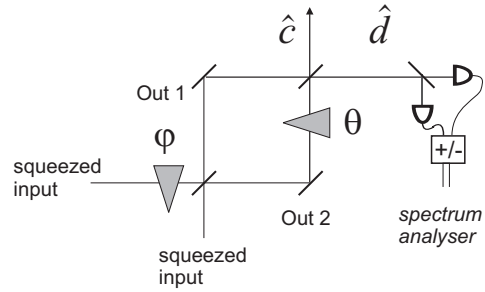}
}
\caption{Set up for quantum interferometry}
\label{interfer}       
\end{center}
\end{figure}
\begin{figure}
\begin{center}
\resizebox{0.35\textwidth}{!}{%
  \includegraphics{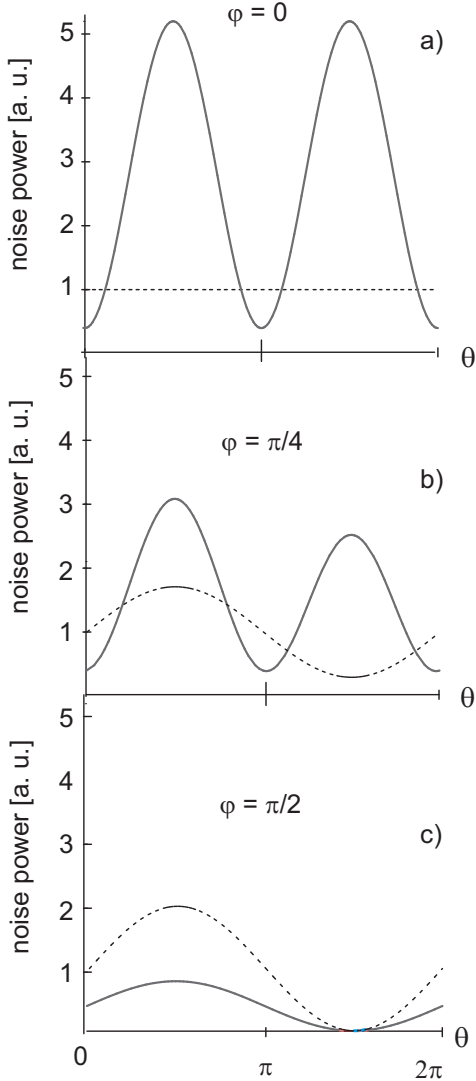}
}
\caption{Intensity noise variances (solid line) and corresponding
shot noise levels (dashed) at the output of the interferometer in
Figure \ref{interfer} versus variable phase shift $\theta$ for
different values of $\varphi$. The curves are plotted for $4$ dB
input squeezing (see Ref. \cite{Silberhorn01}).}
\label{phase}       
\end{center}
\end{figure}

Consider the operation of the setup depicted in Figure
\ref{interfer} in more detail \cite{silberhornLasFoc}. Compared to
the setups in Figures \ref{scheme} and \ref{peres}, the scheme of
Fig. \ref{interfer} in addition allows the phase $\varphi$ of the
first interference to be controlled. That means that the degree of
entanglement between the output beams $1$ and $2$ is tunable. The
noise variances and mean value of the signal at the output of the
interferometer are recorded by balanced detection in at least one
of the output ports (see also Fig. \ref{peres}). From these
measurements the noise variance of the signal normalized to its
shot noise level can be inferred: $V(\delta X_{c,d})= \langle
\left ( \delta X_{\rm c,d} \right )^2 \rangle / V( \delta X_{\rm
c,d, SN})$. In this way the noise reduction at the output of the
interferometer can be evaluated with respect to the interference
phases $\varphi$ and $\theta$. This is illustrated in Figure
\ref{phase}.

For the phase of the first interference equal to zero,
$\varphi=0$, no entanglement is generated. In this case the setup
(Fig. \ref{interfer}) represents just a trivial Mach-Zehnder
interferometer with two squeezed inputs transformed to two
squeezed outputs $\hat c, \hat d$, if the phase of the second
interference is adjusted properly (Fig. \ref{phase} (a)). The
phase relation equal to $\varphi=\frac{\pi}{2}$ corresponds to the
maximal available entanglement between beams $1$ and $2$ (Fig.
\ref{phase} (c)). This is the situation described in the previous
section (Fig. \ref{peres}). The dependence of the second
interference on the phase is given then by equation
(\ref{phaseIndep}). Note that noise reduction observed for
$\varphi=\frac{\pi}{2}$ is insensitive to the phase $\theta$: the
ratio $V(\delta X_{c,d})= \langle \left ( \delta X_{\rm c,d}
\right )^2 \rangle / V( \delta X_{\rm c,d, SN}) < 1$ remains
constant and less than unity for all values of $\theta$ (Fig.
\ref{phase} (c)). Figure \ref{phase} (b) shows the intermediate
situation with a close to maximal degree of entanglement (see also
\cite{Silberhorn01}).

 What do all these noise variances have to do
with quantum interferometry? The proper analysis of the curves
gives an important hint: the main mechanism enabling the
suppression of the resolution limit below that of Eq. (\ref{sql})
is the quantum correlation, i.e. entanglement, between beams $1$
and $2$. To make this statement comprehensive, let us discuss
Figure \ref{phase} again in view of quantum interferometry.

Recall that the limit for minimal resolvable phase modulation can
be derived in terms of signal-to-noise ratio \cite{MilburnWalls}:
\begin{eqnarray}
\displaystyle {\rm SNR} =\frac{ \langle \hat n \rangle}{
\sqrt{V(n)}}=  \frac{ \langle \hat X_{c,d} \rangle}{
\sqrt{V(\delta X_{c,d} )}}\label{snr}
\end{eqnarray}
where $\hat n=\hat c^\dagger \hat c$ is the photon number operator
in mode $\hat c$ with the mean value $\langle \hat n \rangle = n $
(analogously for $\hat d$). The smallest detectable phase shift is
defined to be that for which ${\rm SNR}=1$.

The analysis of  Figure \ref{phase} on this basis yields the
following results. For the phase relation $\varphi=0$ (Fig.
\ref{phase} (a)) there is no interferometric performance of the
setup as the signal is zero. For Figure \ref{phase} (b) minimal
resolvable phase modulations around $\theta = 0 + k \pi$ are:
\begin{eqnarray}
\displaystyle \delta \theta_{\rm min} = \sqrt{\frac{2V_{\rm sq}^+
(\delta X)}{n}}
\end{eqnarray}
where $V_{\rm sq}^+ (\delta X)=V(\delta X)$ is the squeezing
variance determining the maximal available quality of entanglement
for the input squeezed beams with $V(\delta X)$. For the situation
corresponding to maximal available entanglement between the beams
(Fig. \ref{phase} (c)) the minimal resolvable phase modulations
around $\theta = {\pi \over 2}  + k \pi$ drop out as:
\begin{eqnarray}
\displaystyle \delta \theta_{\rm min} = \sqrt{\frac{V_{\rm sq}^+
(\delta X)}{n}}.
\end{eqnarray}
The setup thus reaches the best performance for the best
entanglement.

\begin{figure}
\begin{center}
\resizebox{0.4\textwidth}{!}{%
  \includegraphics{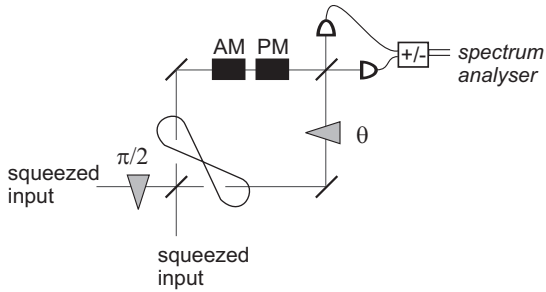}
}
\caption{Setup for quantum dense coding with bright entangled
beams.}
\label{dense}       
\end{center}
\end{figure}
First experiments on quantum interferometry were performed using
squeezed vacuum at the second normally not used input of a
Mach-Zehnder or a Michelson interferometer
\cite{XiaoKimble,Yamamoto}. The important requirement for reaching
sub-shot noise operation is that the squeezed quadrature of the
second input port is $\frac{\pi}{2}$
 out of phase with the coherent excitation of the first input
 port. If one has two intense beams at each input port this
 condition can still be fulfilled \cite{Bjork}.
 The scheme considered in Fig. \ref{interfer} is closely
 related to quantum interferometry
 with two Fock states \cite{HollandBurnett} and to other
interferometric setups used before
\cite{MilburnWalls,Bjork,HollandBurnett}, i.e. it uses two
$\varphi = \pi/2 $ shifted bright squeezed beams as input modes.
However, it possess several novel features. First, noise reduction
performance of the setup is insensitive to the phase $\theta$ for
symmetric beams (Fig. \ref{phase} (c)). Second, so far the ability
to suppress the quantum limit of the interferometer was always
related to  reduced noise in the input fields. Here it is clearly
demonstrated that it is the quantum correlations, quantum
entanglement, between the interferometer arms which is responsible
for the enhanced resolution (see also \cite{Kim}). Third, it is
not necessary to go for the detection of the difference signal of
two outputs $\hat c^\dagger \hat c- \hat d^\dagger \hat d$ because
the measurement in one arm provides the same precision. Fourth, as
the noise reduction is phase insensitive, one can choose an
arbitrary point of operation $\theta \neq \pi/2, 0$ which might be
advantageous in a particular application.

\subsection{Quantum dense coding}
The phase-insensitive noise reduction performance of the
interferometric scheme (Fig. \ref{interfer}) makes it a promising
setup for dense coding \cite{DenseBennett,DenseBraunstein}.
Quantum dense coding was first suggested  \cite{DenseBennett} and
realized in an experiment \cite{DenseMattle} for discrete quantum
variables like polarization states of entangled photon pairs. In
this case it aims to enhance the classical information capacity of
quantum communication channels beyond the Holevo's bound
\cite{Holevo}.  Recently the idea of quantum dense coding was
extended to the quadrature components of the electromagnetic field
using entangled light beams as quantum channel
\cite{DenseBraunstein}. For such continuous variables, dense
coding refers to the ability to read out amplitude and phase
modulations with a precision below the limit given by the
Heisenberg uncertainty relation. For large photon numbers, the
channel capacity of the scheme \cite{DenseBraunstein} approaches
twice that of classical coherent-state communication
\cite{DenseBraunstein}. The first experimental implementation of
quantum dense coding with bright entangled beams was proposed in
Ref. \cite{DensePeng} and recently implemented  \cite{CLEOPeng}.

Figure \ref{dense} shows the interferometric setup (Fig.
\ref{interfer}) modified to implement dense coding using intense
entangled beams. This corresponds to the setup shown in Ref.
\cite{DensePeng} where this scheme was  only briefly addressed.
The analysis was based on the squeezing characteristics of the
input beams for entanglement generation. Here we would like to
emphasize the role of the quantum correlations $V_{\rm sq}^+
(\delta X)$, $V_{\rm sq}^- (\delta Y)$  on the scheme performance,
which means on being able to read out amplitude and phase
modulations with a precision below the Heisenberg limit. We assume
$V_{\rm sq}^+ (\delta X) = V_{\rm sq}^- (\delta Y) = V_{\rm sq}$
which is required for the phase independent performance of the
scheme. The balanced detection of the combined mode of both
outputs delivers the information about small imposed amplitude
$\delta X_{\rm m}$ and phase $\delta Y_{\rm m}$ modulations:
\begin{eqnarray}
V^+ = \alpha ^2 \left ( V_{\rm sq} + V (\delta X_{\rm m}) \right
), \label{code1}
\\
V^- = \alpha ^2 \left ( V_{\rm sq} + V (\delta Y_{\rm m}) \right
)\label{code2}
\end{eqnarray}
where $V^\pm$ are the noise variances recorded in the sum and
difference detector channels, respectively. The Heisenberg
relation imposes a lower resolution bound for coherent
communication and  there will be $V_{\rm coh} = 1$ in the place of
$V_{\rm sq}$ in equations (\ref{code1}, \ref{code2}). Quantum
EPR-like correlations between the bright beams in the two
interferometer arms imply $V_{\rm sq} \ll 1$.  Hence the use of
quantum entanglement enhances  the resolution of quadrature
modulations beyond the Heisenberg limit for all values of $\theta$
(\ref{code1}, \ref{code2}). This enables quantum dense coding
independent of the interference phase $\theta$. The
phase-insensitivity eliminates the need for extensive setup
stabilization. It makes the scheme more reliable and easier to
handle, in particular for the communication over long distances.

\section{Conclusions}
The generation and experimental characterization of bright beam
entanglement has been reviewed. We report on an entanglement-based
interferometric setup using intense optical fields (Fig.
\ref{peres}, \ref{interfer}, \ref{dense}). This setup can be
modified to implement various optical techniques ranging from
entanglement evaluation through quantum interferometry and dense
coding to quantum teleportation \cite{BraunsteinKimble,Leuchs99}.
In all these applications only direct detection is required
without the need for local oscillator fields. The role of quantum
EPR-like correlations, $V_{\rm sq}$, was explicitly shown, as well
as phase-insensitive performance of the schemes (Fig. \ref{peres},
\ref{interfer}, \ref{dense}) provided that there is equal
squeezing in the input beams.

A  phase-insensitive scheme for quantum interferometry with bright
entangled beams is proposed (Fig.  \ref{interfer}). The resolution
limit for phase modulations can be suppressed up to $\displaystyle
\sqrt{\frac{V_{\rm sq}(\delta X)}{n}}$ as compared to
$\displaystyle \sqrt{\frac{1}{n}}$ where $V_{sq} (\delta X)$ is
determined by the quantum amplitude correlations of the entangled
beams.

In the schemes outlined above it was sufficient to infer the
information about the phase quadrature from the noise variances
obtained in direct amplitude detection. However, there are a
number of quantum information protocols which require the explicit
measurement of all involved conjugate quantum variables, for
example, quantum key distribution with continuous variables
(\cite{Lorenz,SilberhornQKD} and references therein). Continuous
variable polarization entanglement \cite{Korolkova01} can be
extremely useful for such purposes. For this type of entanglement
all the relevant variables can be detected without the need for
auxillary local oscillator fields or other evolved measurement
techniques.

This work was supported by the Deutsche Forschungsgemeinschaft and
by the EU grant under QIPC, project IST-1999-13071 (QUICOV). The
authors gratefully acknowledge the help of J. Heersink in the
preparation of the manuscript.

\end{document}